\documentclass[12pt,a4paper]{article}
\usepackage{a4wide,cite}

\newcommand{\be}{\begin{equation}}
\newcommand{\ee}{\end{equation}}
\newcommand{\bea}{\begin{eqnarray}}
\newcommand{\eea}{\end{eqnarray}}
\newcommand{\ep}{\varepsilon}

\newcommand{\nn}{\nonumber}
\newcommand{\sfrac}[2]{{\textstyle{{#1}\over{#2}}}}

\begin{document}

\thispagestyle{empty}
\begin{flushright}
{MZ-TH/98-49} \\[5mm]
{hep-ph/9811503} \\[5mm]
{November 1998}           \\
\end{flushright}
\vspace*{3cm}
\begin{center}
{\bf \Large
Two-loop \ QCD \ vertices, \ WST \ identities \ 
and \ RG \ quantities\footnote{Talk presented by A.~D.
at the Workshop on Exact Renormalization Group
(Faro, Portugal, September 1998).
To be published in the proccedings (World Scientific). }}
 \end{center}
\vspace{1cm}
\begin{center}
A.~I.~Davydychev$^{a,}$\footnote{davyd@thep.physik.uni-mainz.de .
                                 On leave from
                                 Institute for Nuclear Physics,
                                 Moscow State University,
                                 119899, Moscow, Russia.},
P. Osland$^{b,}$\footnote{Per.Osland@fi.uib.no}
 \ \ and \ \
O. V. Tarasov$^{c,}$\footnote{tarasov@itp.unibe.ch .
                                 On leave from
                                 Joint Institute for Nuclear Research,
                                 141980, Dubna, Russia}
\\
 \vspace{1cm}
$^{a}${\em   
Department of Physics, University of Mainz, \\
      Staudingerweg 7, D-55099 Mainz, Germany}
\\
\vspace{.3cm}
$^{b}${\em
Department of Physics, University of Bergen, \\
      All\'{e}gaten 55, N-5007 Bergen, Norway}
\\
\vspace{.3cm}
$^{c}${\em
Institute of Theoretical Physics, University of Bern, \\
          Sidlerstrasse 5, CH-3008, Bern, Switzerland}
\end{center} 
 \hspace{3in}
 \begin{abstract}
Recent results from the study of the two-loop 
three-gluon and ghost-gluon vertices are reviewed.
The relevant Ward--Slavnov--Taylor (WST) identities
and Renormalization Group (RG) quantities are discussed.
 \end{abstract}

\newpage

%#####################################################################
\section{Introduction}
%#####################################################################

The one-loop QCD vertices have been known for quite some time.
In the symmetric case, $p_1^2=p_2^2=p_3^3$,
the one-loop result for the three-gluon vertex
in an arbitrary covariant gauge was calculated
by Celmaster and Gonsalves~\cite{CG} and confirmed
by Pascual and Tarrach~\cite{PT}.
The general off-shell case was considered in the Feynman gauge by
Ball and Chiu~\cite{BC2}.
Later, various on-shell results have also been given,
by Brandt and Frenkel~\cite{BF},
and by Nowak, Prasza{\l}owicz and S{\l}omi{\'n}ski~\cite{NPS}.
The most general results, valid for arbitrary values
of the space-time dimension and the covariant-gauge 
parameter, have been presented in our paper~\cite{DOT1}.
The one-loop quark-gluon vertex (or its 
Abelian part which is related to the QED vertex) 
was studied by several authors~\cite{qqg}.

Two-loop corrections to the three-gluon and ghost-gluon vertices 
have been studied in the {\em zero-momentum limit},
which refers to the case when one gluon
has vanishing momentum.
In this limit, the renormalized expressions for QCD vertices
in the Feynman gauge have been presented by Braaten and
Leveille~\cite{BL}.
In an arbitrary covariant gauge, the relevant results have been
presented in our previous paper~\cite{DOT2};
they will be reviewed here. 
It should be noted that the zero-momentum limit of the 
three-gluon vertex,
as well as the relevant limits of the ghost-gluon vertex, are
infrared finite, i.e. we do not get any singularities of
infrared (on-shell) origin. The main argument~\cite{DOT2} is 
just power counting.

Another limit of interest corresponds to an on-shell configuration, 
when two external gluons are on the mass shell,
$p_1^2=p_2^2=0$. This limit possesses essential infrared (on-shell)
singularities~\cite{onshell}.
The relevant on-shell results for the three-gluon and 
ghost-gluon vertices have been recently presented
by two of us~\cite{DO1}.

Information about Green functions is also required
for the calculation of certain quantities related to the
renormalization group equations, such as the $\beta$ function
and anomalous dimensions. The two-loop-order contributions 
to these quantities were calculated
by Caswell~\cite{Caswell}, Jones~\cite{Jones},
Vladimirov and Tarasov~\cite{VlaTar}, and Egorian and
Tarasov~\cite{EgTar}, whereas the 
three-loop-order results were obtained by 
Tarasov, Vladimirov and Zharkov \cite{TVZh},
and by Larin and Vermaseren~\cite{LV}. 
Moreover, recently the four-loop-order expressions became
available, due to Larin, van Ritbergen and Vermaseren~\cite{LvRV}.

%#################################################################
\section{Preliminaries}
%#################################################################

The lowest-order gluon propagator in  a general covariant gauge is
\begin{equation}
\label{gl_prop}
\left(\delta^{a_1 a_2}/p^2\right)
\left( g_{\mu_1 \mu_2} - \xi_B \; p_{\mu_1} p_{\mu_2}/p^2
\right) ,
\end{equation}
where $\xi_B\equiv1-\alpha_B$ is the {\em bare} gauge parameter.

The three-gluon vertex is defined as
\begin{equation}
\label{ggg}
\Gamma_{\mu_1 \mu_2 \mu_3}^{a_1 a_2 a_3}(p_1, p_2, p_3)
\equiv  - \mbox{i} \; g_B \;
f^{a_1 a_2 a_3} \; \Gamma_{\mu_1 \mu_2 \mu_3}(p_1, p_2, p_3) ,
\end{equation}
where $f^{a_1 a_2 a_3}$
are the totally antisymmetric colour structures corresponding
to the adjoint representation of the gauge group
(for example, $\mbox{SU}(N)$); $g_B$ is the {\em bare} 
coupling constant.
When one of the momenta is zero, the three-gluon vertex
contains only two tensor structures~\cite{BL},
\bea
\label{BL-decomp}
\Gamma_{\mu_1 \mu_2 \mu_3}(p, -p, 0)
= \left( 2 g_{\mu_1 \mu_2} p_{\mu_3} - g_{\mu_1 \mu_3} p_{\mu_2}
	 - g_{\mu_2 \mu_3} p_{\mu_1} \right)  T_1(p^2)
\nn \\
- p_{\mu_3} \left( g_{\mu_1 \mu_2}
		      - p_{\mu_1} p_{\mu_2}/p^2 \right) 
					       T_2(p^2) .
\eea
At the lowest, ``zero-loop'' order, 
we have $T_1^{(0)}=1, \; T_2^{(0)}=0$.

The ghost-gluon vertex can be represented as
\begin{equation}
\label{ghg}
\widetilde{\Gamma}_{\mu_3}^{a_1 a_2 a_3}(p_1, p_2; p_3)
\equiv -\mbox{i} g_B \; f^{a_1 a_2 a_3} \;
{p_1}^{\mu} \; \widetilde{\Gamma}_{\mu \mu_3}(p_1, p_2; p_3) ,
\end{equation}
where $p_1$ is the out-ghost momentum, $p_2$ is the in-ghost momentum,
$p_3$ and $\mu_3$ are the momentum and the Lorentz index of the gluon
(all momenta are ingoing). For $\widetilde{\Gamma}_{\mu \mu_3}$, 
the following decomposition~\cite{BC2} is useful:
\be
\label{BC-ghg}
\widetilde{\Gamma}_{\mu \mu_3}(p_1,p_2;p_3)
= g_{\mu \mu_3} a(p_3,p_2,p_1)
+ \left\{ \mbox{terms with}\;\; {p_i}_{\mu} {p_j}_{\mu_3} \right\} .
\ee
At the ``zero-loop'' level,
$\widetilde{\Gamma}^{(0)}_{\mu \mu_3} = g_{\mu \mu_3}$, $a^{(0)}=1$.

In particular, when $p_3=0$ or $p_2=0$ we get
\bea
\label{ghg1}
\widetilde{\Gamma}_{\mu \mu_3}(-p,p;0)
= g_{\mu \mu_3} a_3(p^2) 
+ \left\{ p_{\mu} p_{\mu_3} \mbox{term} \right\},
\nn \\
\widetilde{\Gamma}_{\mu \mu_3}(p,0;-p)
= g_{\mu \mu_3} a_2(p^2)
+ \left\{ p_{\mu} p_{\mu_3} \mbox{term} \right\},
\nn \\
a_3(p^2)\equiv a(0,p,-p), \hspace{4mm}
a_2(p^2)\equiv a(-p,0,p) .
\eea

The gluon polarization operator 
and the ghost self energy are defined as
\begin{equation}
\label{gl_po}
\Pi_{\mu_1 \mu_2}^{a_1 a_2}(p)
\equiv - \delta^{a_1 a_2}
\left( p^2 g_{\mu_1 \mu_2} - {p}_{\mu_1}{p}_{\mu_2} \right) J(p^2),
\end{equation}
\begin{equation}
\label{gh_se}
\widetilde{\Pi}^{a_1 a_2}(p^2) = \delta^{a_1 a_2} \; p^2 \;
\left[G(p^2)\right]^{-1} .
\end{equation}
In the lowest-order approximation $J^{(0)}=G^{(0)}=1$.

We shall use dimensional regularization~\cite{dimreg},
with the space-time dimension $n=4-2\ep$, $\ep\to0$.
In this paper we adopt the modification of the
renormalization prescription by `t~Hooft~\cite{Hooft},
corresponding to the so-called $\overline{\mbox{MS}}$
scheme~\cite{MSbar}.
In what follows, the notations
$\xi$, $\alpha$, $g$ (without subscript) correspond
to the {\em renormalized} (in the $\overline{\mbox{MS}}$ scheme)
quantities.   
In eqs.~(\ref{gl_prop}), (\ref{ggg}), (\ref{ghg})
they are understood
as the {\em bare} quantities $\xi_B$, $\alpha_B$, $g_B$.

The renormalization
constants $Z_{\Gamma}$ relating the dimensionally-regularized
one-particle-irreducible Green functions to the renormalized   
ones,
\begin{equation}
\label{renormalization}
\Gamma^{{\rm (ren)}}\left(\left\{\frac{p_i^2}{\mu^2}\right\},
\alpha,g^2\right)=
\lim_{\varepsilon \to 0}
\left[
Z_{\Gamma}\left(\frac{1}{\varepsilon},\alpha,g^2\right)
\Gamma\left(\{p_i^2\},\alpha_B,g^2_B,\varepsilon\right)
\right],
\end{equation}
look in this scheme like
\begin{equation}
\label{Z_Gamma}
Z_{\Gamma}\left( \frac{1}{\varepsilon},\alpha,g^2
         \right)=1+\sum_{j=1}^{\infty}
C_{\Gamma}^{[j]}(\alpha,g^2)  \frac{1}{\varepsilon^j},
\end{equation}
where $\alpha=1-\xi$.
In eq.~(\ref{renormalization}) $\mu$ is the renormalization
parameter with the dimension of mass.
It is assumed that on the r.h.s.
of eq.~(\ref{renormalization}) the squared bare charge
$g_B^2$ and the bare gauge parameter $\alpha_B$
must be substituted in terms of renormalized ones,
multiplied by appropriate $Z$ factors (cf. eqs.~(\ref{g_Z})
and (\ref{alpha_Z})).

We use the following definitions for renormalization factors:
\begin{eqnarray}
\label{defZ1}
&&\Gamma_{\mu_1 \mu_2 \mu_3 }^{{\rm (ren)}}(p_1,p_2,p_3)
 =Z_1\;\Gamma_{\mu_1 \mu_2 \mu_3 }(p_1,p_2,p_3),
\\
\label{defZ1tilde}
&&
\Pi^{{\rm (ren)}\;a_1 a_2}_{\mu_1 \mu_2}(p)
   =Z_3 \;\Pi^{a_1 a_2}_{\mu_1 \mu_2}(p),
\\
\label{defZ3}
&&\widetilde{\Gamma}_{\mu}^{{\rm (ren)}\; a_1 a_2 a_3}(p_1,p_2,p_3)
=\widetilde{Z}_1 \; \tilde{\Gamma}_{\mu}^{a_1 a_2 a_3}(p_1,p_2,p_3),
\\
\label{defZ3tilde}
&&\widetilde{\Pi}^{{\rm (ren)}\; a_1 a_2}(p^2)
 =\widetilde{Z}_3 \; \widetilde{\Pi}^{a_1 a_2}(p^2) .
\end{eqnarray}
The WST identity requires that
\be
\label{WST-Z}
Z_3/Z_1=\widetilde{Z}_3/\widetilde{Z}_1 .
\ee

The results for these renormalization factors 
in the pure Yang--Mills theory were  
first presented by Jones~\cite{Jones} (Feynman gauge) and 
by Vladimirov and Tarasov~\cite{VlaTar} (an arbitrary covariant gauge).
The complete results in an arbitrary covariant gauge, including the
fermionic contributions, were presented 
by Egorian and Tarasov~\cite{EgTar}
(cf.\ also in refs.~\cite{PT-QCD,DOT2}).

Using (\ref{WST-Z}), the bare coupling constant $g_B^2$
can be chosen (in the $\overline{\mbox{MS}}$ scheme) as
\begin{equation}
\label{g_Z}
g_B^2=
\left[\mu^2 e^{\gamma}/(4\pi)\right]^{\ep}
g^2 \widetilde{Z}_1^2
Z_3^{-1}\widetilde{Z}_3^{-2}
=\left[\mu^2 e^{\gamma}/(4\pi)\right]^{\ep}
g^2 Z_1^2 Z_3^{-3},
\end{equation}
where $\gamma$ is the Euler constant.
The gauge parameter $\alpha=1\!-\!\xi$ is renormalized as
\be
\label{alpha_Z}
\alpha_B=Z_3 \alpha,
\hspace{10mm} \mbox{so that} \hspace{10mm}
\xi_B = 1- Z_3 (1-\xi) .
\ee

Below we shall also use the notation
\be
h \equiv g^2/(4\pi)^2 = \alpha_s/(4\pi) ,
\hspace{10mm}
\mbox{where}
\hspace{10mm}
\alpha_s\equiv g^2/(4\pi) .
\ee

%####################################################################
\section{WST identities}
%####################################################################

In a covariant gauge, the Ward--Slavnov--Taylor (WST) 
identity~\cite{WST}
for the three-gluon vertex is of the following form~\cite{MarPag}:
\begin{eqnarray}
\label{WST}
p_3^{\mu_3} \Gamma_{\mu_1 \mu_2 \mu_3}(p_1, p_2, p_3)
= - J(p_1^2) G(p_3^2)
\left( g_{\mu_1 \;\;\;}^{\;\;\; \mu_3}\; p_1^2
       - {p_1}_{\mu_1} \; {p_1}^{\mu_3} \right) 
\widetilde{\Gamma}_{\mu_3 \mu_2}(p_1, p_3; p_2)
\nonumber \\
 + J(p_2^2) G(p_3^2) 
\left( g_{\mu_2 \;\;\;}^{\;\;\; \mu_3}\; p_2^2
       - {p_2}_{\mu_2} \; {p_2}^{\mu_3} \right) 
\widetilde{\Gamma}_{\mu_3 \mu_1}(p_2, p_3; p_1) .
\hspace*{3mm}
\end{eqnarray}

Consider what follows from (\ref{WST}) in the limit when one of the
momenta vanishes.
Contracting with a non-zero momentum, we get 
\begin{equation}
\label{WST-ord}
p^{\mu_1} \Gamma_{\mu_1 \mu_2 \mu_3}(p, -p, 0)
= - J(p^2) G(p^2) a_3(p^2)
\left( g_{\mu_2 \mu_3} p^2 - p_{\mu_2} p_{\mu_3} \right) ,
\end{equation}
where $a_3(p^2)$ is defined in eq.~(\ref{ghg1}). 
Considering contraction with the vanishing momentum,
we get a {\em differential} WST identity.
It can be written in a way
which involves just the $a$ functions from the ghost-gluon 
vertex~\cite{DOT2}.
For the scalar functions $T_1(p^2)$ and $T_2(p^2)$, the differential 
WST identity reads
\be
\label{T1-WST}
T_1(p^2)=a_3(p^2) \; G(p^2) \; J(p^2) ,
\ee
\be
\label{T2-WST}
T_2(p^2) \!=\! 2 T_1(p^2) 
\!-\! 2 G(0) \!
\left[ 
a_2(p^2) \frac{\mbox{d}}{\mbox{d} p^2}\!\left( p^2 J(p^2) \right)
\!-\! p^2 J(p^2) \frac{\mbox{d}a_2(p^2)}{\mbox{d} p^2}
\!+\! \widetilde{a}_2(p^2) J(p^2)
\right],
\ee
where the function $\widetilde{a}_2(p^2)$ is defined as
\begin{equation}
\label{a_tilde_2}
\widetilde{a}_2(p^2) \equiv
\left. {p_1}_{\sigma}\; \frac{\partial}{\partial {p_1}_{\sigma}} \;
          a(p_3,-p_1-p_3,p_1)\right|_{p_1=-p_3=p} \; \; .
\end{equation}   
It can be calculated directly at the diagrammatic level (see below).

Therefore, the differential WST identity  makes it possible to 
define the whole
three-gluon vertex (not only its longitudinal part)
in terms of two-point functions and the ghost-gluon vertex.
Moreover, it can be used as another independent way, in addition to
the direct calculation, to obtain results for the three-gluon
vertex.

%#####################################################################
\section{Results for the three-gluon vertex}
%#####################################################################

The results for unrenormalized one-loop contributions 
to the scalar functions 
$T_1(p^2)$ and $T_2(p^2)$ (in arbitrary space-time dimension)
can be found in ref.~\cite{DOT1}, eqs.~(4.30), (4.31), (4.33)
and (4.34).

The diagrams contributing to the three-gluon vertex at the two-loop
level are shown in Fig.~1 of ref.~\cite{DOT2}.
Note that the non-planar diagrams do not contribute,
since their colour structures vanish~\cite{Cvit}.
When one external momentum vanishes, technically the problem 
reduces to the calculation of two-point two-loop Feynman
integrals. To calculate the occurring integrals with higher
powers of the propagators, the integration-by-parts 
procedure~\cite{ibp} has been used. For the integrals with numerators,
some other known algorithms~\cite{ibp,PLB'91} were employed.
Straightforward calculation of the sum
of all these contributions yields the results for
the unrenormalized scalar functions
which are presented in eqs.~(4.8)--(4.11) of ref.~\cite{DOT2}.
 
Using eq.~(\ref{defZ1}) with
\bea
\label{Z1}
Z_1=1
+\frac{h}{\ep} \left[ C_A\left(\sfrac{2}{3}+\sfrac{3}{4}\xi\right)
                      -\sfrac{4}{3}T \right]
+h^2 \left\{
C_A T \left[ \frac{1}{\ep^2}\left(\sfrac{5}{2}-\xi\right)
-\frac{25}{12\ep} \right]
 - \frac{2}{\ep}C_F T
\right.
\nn \\
\left.
+C_A^2\left[ \frac{1}{\ep^2}
\left(-\sfrac{13}{8}-\sfrac{7}{16}\xi+\sfrac{15}{32}\xi^2\right)
+\frac{1}{\ep}
\left(\sfrac{71}{48}+\sfrac{45}{32}\xi-\sfrac{3}{16}\xi^2\right)
\right]
\right\} + {\cal{O}}(h^3),
\hspace*{3mm}
\eea
we obtain the renormalized
scalar amplitudes appearing in the three-gluon vertex
(cf.\ eq.~(\ref{BL-decomp})),
\bea
\label{T1ren}
T_1^{{\rm (ren)}} =1
 +h \left[
 C_A\left(-\sfrac{35}{18}+\sfrac{1}{2}\xi-\sfrac{1}{4}\xi^2\right)
 + \sfrac{20}{9} T
    \right]  
\hspace{50mm}
\nn \\
+h^2 \left[
 C_A^2  \left(
 - \sfrac{4021}{288} - \sfrac{1}{4} \zeta_3
 -\sfrac{2317}{576} \xi  + \sfrac{15}{8}\xi \zeta_3
 + \sfrac{113}{144} \xi^2  -\sfrac{1}{16} \xi^3
 + \sfrac{1}{16} \xi^4  \right)  
\right.
\hspace{3mm}
\nn \\
\left.
 + C_A T  \left( \sfrac{875}{72}  + 8 \zeta_3 + \sfrac{20}{9}\xi  
       -\sfrac{10}{9} \xi^2 \right)
+ C_F T \left( \sfrac{55}{3} - 16 \zeta_3 \right)
\right] + {\cal{O}}(h^3),
\hspace{3mm}
\eea
\bea
\label{T2ren}
T_2^{{\rm (ren)}}=  
h\left[   
C_A \left(- \sfrac{4}{3} - 2\xi\! + \sfrac{1}{4} \xi^2 \right)
\!+ \sfrac{8}{3} T \right]
+h^2 \left[ C_A T \left(
 \sfrac{157}{9}-\sfrac{37}{18}\xi\!-\sfrac{2}{9} \xi^2 \right)
\!+8 C_F T
 \right.   
\nn \\
\left.
+ C_A^2 \left(
-\sfrac{641}{36}- \zeta_3
+ \sfrac{5}{18} \xi-\sfrac{1}{2} \xi \zeta_3
-\sfrac{287}{144} \xi^2 + \sfrac{19}{16} \xi^3
 - \sfrac{1}{8} \xi^4
\right) \right]
+ {\cal{O}}(h^3). 
\hspace*{4mm}
\eea
Here and henceforth,
we put $p^2=-\mu^2$ in the renormalized expressions.
In Feynman gauge our expressions agree with
those by Braaten and Leveille~\cite{BL}.

In eqs.~(\ref{T1ren}) and (\ref{T2ren})
we use the standard notation $C_A$ for the eigenvalue
of the quadratic Casimir operator in the {\em adjoint} representation,
$f^{acd}f^{bcd} = C_A\delta^{ab}$ ($C_A = N$ for the SU($N$) group),
whereas
$C_F$ is the eigenvalue of this operator
in the {\em fundamental} representation
($C_F=(N^2-1)/(2N)$ for the $\mbox{SU}(N)$ group).
Furthermore,
$T\equiv N_f T_R$, 
$T_R = {\textstyle{1\over8}} \; \mbox{Tr}(I)={\textstyle{1\over2}}$,
where $I$ is the ``unity'' in the space of Dirac matrices,
$N_f$ is the number of quarks,
$\zeta_3\equiv\zeta(3)$
is the value of Riemann's zeta function.

%##################################################################
\section{Results for the ghost-gluon vertex}
%##################################################################

In order to check the WST identity, we need results for the
ghost-gluon vertex in two limits corresponding to eq.~(\ref{ghg1}). 
We shall also need the derivative 
$\widetilde{a}_2(p^2)$, eq.~(\ref{a_tilde_2}). 
The relevant one-loop results (for an arbitrary $n$)
are listed in Appendix~A of ref.~\cite{DOT2}. 
Two-loop contributions to the ghost-gluon vertex are shown in Fig.~2
of ref.~\cite{DOT2}.
Straightforward calculation gives the unrenormalized results
presented in eqs.~(5.6)--(5.13) of ref.~\cite{DOT2}.

The derivative (\ref{a_tilde_2}) has been
calculated in the following way~\cite{DOT2}.
Let us consider the momenta $p_1$ and $p_3$
as independent variables, whereas
$p_2=-p_1-p_3$. Therefore, the momentum $p_1$ flows from
the in-ghost leg to the out-ghost leg. An unambiguous
$p_1$ path inside the diagram can be chosen as the one
coinciding with the ghost line. This is convenient,
since all we need to differentiate are just two types of objects: 
ghost propagators and ghost-gluon vertices occurring along this path.
In this way, we avoid differentiating gluon propagators
and three-gluon vertices. We also avoid getting third powers
of propagators.
Technically,
the propagators and vertices along the ghost path were
``marked'' by introducing an extra argument (say, $z$).
Then, the derivative with respect to $z$ was considered,
and the rules for differentiating the ghost-gluon vertex
and the ghost propagator (with subsequent contraction
with ${p_1}_{\mu_1}$) were supplied.
In this way, we just formally differentiate along the 
ghost line, and then perform all calculations for $p_1=-p_3=p$,
$p_2=0$. Finally, extracting the coefficient of $g_{\mu \mu_3}$ 
we arrived at the results for the function (\ref{a_tilde_2})
presented in eqs.~(5.14), (5.15) of ref.~\cite{DOT2}.

Using eq.~(\ref{defZ1tilde}) with
\be
\label{Z1tilde}
\widetilde{Z}_1=1-\frac{h}{2\ep} C_A (1-\xi)
+h^2 C_A^2 (1-\xi)
\left[
\frac{1}{\ep^2} \left(\sfrac{5}{8} - \sfrac{1}{4}\xi \right)
+ \frac{1}{\ep} \left( -\sfrac{3}{8}+\sfrac{1}{16}\xi \right)
\right]
+ {\cal{O}}(h^3),
\ee
we obtain
the renormalized expressions for the scalar functions
occurring in the ghost-gluon vertex: 
\bea
a_3^{{\rm (ren)}}
=1 + \sfrac{1}{2} \; h \; C_A \; (1-\xi)
\hspace{91mm}
\nn \\
+ h^2 \left[ C_A^2
\left( \sfrac{137}{48} - \sfrac{1}{2} \zeta_3\! - \sfrac{299}{96}\xi
      -\sfrac{1}{8}\xi\zeta_3\! + \sfrac{7}{16}\xi^2
      + \sfrac{3}{16}\xi^2\zeta_3 \right) + \sfrac{1}{4} C_A T \right]
+ {\cal{O}}(h^3),
\hspace*{3mm}
\eea
\bea
a_2^{{\rm (ren)}}
=1+ \sfrac{1}{4} \; h \; C_A \; \xi (1-\xi)
\hspace{92mm}
\nn \\
+ h^2 (1\!-\!\xi) \left[ C_A^2
\left( \sfrac{167}{72} - \zeta_3\! - \sfrac{43}{144}\xi
       - \sfrac{1}{16}\xi^2
       - \sfrac{1}{16}\xi^3 \right)
- \sfrac{5}{9} C_A T (1\!-\!\xi) \right]
+ {\cal{O}}(h^3) .
\hspace*{3mm}
\eea

%#################################################################
\section{Results for the two-point functions}
%#################################################################

One-loop results in arbitrary space-time dimension are available
e.g.\ in refs.~\cite{Muta,DOT1} (see also in Appendix~A of 
ref.~\cite{DOT2}).
Calculating the sum of one-particle irreducible
two-loop diagrams contributing to the gluon polarization operator
and the ghost self energy (shown in Fig.~3 and Fig.~4 of 
ref.~\cite{DOT2}, respectively), we arrive at the  
results which are presented in eqs.~(6.10)--(6.15)
of ref.~\cite{DOT2}.

Using eqs.~(\ref{defZ3}) and (\ref{defZ3tilde}) with
\bea
\label{Z3}
Z_3=1  
+\frac{h}{\ep}\left[ C_A \left(\sfrac{5}{3}+\sfrac{1}{2}\xi\right)
-\sfrac{4}{3}T \right]
+h^2\left\{
C_A T \left[ \frac{1}{\ep^2}
\left( \sfrac{5}{3} - \sfrac{2}{3}\xi \right) - \frac{5}{2\ep} \right]
- \frac{2}{\ep} C_F T
\right.
\nn \\
\left.
+C_A^2 \left[ \frac{1}{\ep^2}
\left( -\sfrac{25}{12} + \sfrac{5}{24}\xi + \sfrac{1}{4}\xi^2 \right)
+\frac{1}{\ep}
\left( \sfrac{23}{8} + \sfrac{15}{16}\xi - \sfrac{1}{8}\xi^2 \right)
\right] \right\}
+ {\cal{O}}(h^3),
\hspace*{3mm}
\eea
\bea   
\label{Z3tilde}  
\widetilde{Z}_3=1+\frac{h}{\ep} C_A
\left( \sfrac{1}{2} + \sfrac{1}{4}\xi \right)
+ h^2 \left\{
C_A^2 \left[
\frac{1}{\ep^2}
\left( -1 - \sfrac{3}{16}\xi + \sfrac{3}{32}\xi^2 \right)
+ \frac{1}{\ep}\left(\sfrac{49}{48}-\sfrac{1}{32}\xi\right)
\right]
\right.
\nn \\
\left. 
+ C_A T \left( \frac{1}{2\ep^2} - \frac{5}{12\ep} \right)
\right\}
+ {\cal{O}}(h^3),
\hspace*{10mm}
\eea
we obtain
the renormalized expressions for two-point functions
\bea
\label{Jren}
J^{{\rm (ren)}}=1+h\left[
C_A \left( -\sfrac{31}{9} + \xi - \sfrac{1}{4}\xi^2 \right)
+ \sfrac{20}{9} T \right]
\hspace{54mm}
\nn \\
+ h^2 \left[
C_A^2 \left( -\sfrac{3245}{144}+\zeta_3-\sfrac{287}{96}\xi+2\xi\zeta_3
             +\sfrac{61}{72}\xi^2-\sfrac{3}{16}\xi^3+\sfrac{1}{16}\xi^4
      \right)
\right.
\hspace{12mm}
\nn \\
\left.
+ C_A T \left( \sfrac{451}{36} + 8\zeta_3 + \sfrac{10}{3}\xi
              -\sfrac{10}{9}\xi^2 \right)
+ C_F T \left( \sfrac{55}{3} - 16\zeta_3 \right) \right]
+ {\cal{O}}(h^3),
\hspace*{3mm}
\eea
\be
G^{{\rm (ren)}}=1+h C_A + h^2 \left[
C_A^2 \left( \sfrac{997}{96} - \sfrac{3}{4}\zeta_3\! -\sfrac{41}{64}\xi
             +\sfrac{3}{8}\xi^2 - \sfrac{3}{16} \xi^2 \zeta_3 \right)\!
- \sfrac{95}{24} C_A T \right]
+ {\cal{O}}(h^3).
\ee
In Feynman gauge, 
these results agree with those presented in ref.~\cite{BL}.

%%%%%%%%%%%%%%%%%%%%%%%%%%%%%%%%%%%%%%%%%%%%%%%%%%%%%%%%%%%%%%%%%%%%%%%%
\section{Renormalization group quantities}
%%%%%%%%%%%%%%%%%%%%%%%%%%%%%%%%%%%%%%%%%%%%%%%%%%%%%%%%%%%%%%%%%%%%%%%%

Using the $1/\ep$ term of the renormalization factor $Z_{\Gamma}$
(cf. eq.~(\ref{Z_Gamma})), one can obtain the corresponding
anomalous dimension $\gamma_{\Gamma}$ via 
\be
\label{an_dim}
\gamma_{\Gamma}\left(\alpha,g^2\right)=
g^2 \frac{\partial}{\partial g^2} 
C_{\Gamma}^{[1]}\left(\alpha,g^2\right) .
\ee
We have checked that in the Feynman gauge $\xi=0$ ($\alpha=1$)
the results for the anomalous dimensions $\widetilde{\gamma}_1$,
$\gamma_3$ and $\widetilde{\gamma}_3$ coincide 
(in the two-loop approximation) with those from ref.~\cite{TVZh}:
\begin{eqnarray*}
\gamma_1&=&h\left[ C_A\left(\sfrac{2}{3}\!+\sfrac{3}{4}\xi\right)
\!-\sfrac{4}{3}T\right]\!
+h^2\left[C_A^2\left(\sfrac{71}{24}\!+\sfrac{45}{16}\xi\!
-\sfrac{3}{8}\xi^2\right)\!-\sfrac{25}{6}C_A T\! -\! 4C_F T \right]
%+{\cal{O}}(h^3),
\!+ ... ,
\\
\widetilde{\gamma}_1&=&-\sfrac{1}{2}h C_A (1-\xi)
+h^2 C_A^2 (1-\xi) \left(-\sfrac{3}{4}+\sfrac{1}{8}\xi\right)
+ ... \; ,
%+{\cal{O}}(h^3),
\hspace{36mm}
\\
\gamma_3&=&h\left[C_A\left(\sfrac{5}{3}\!+\sfrac{1}{2}\xi\right)
\!-\sfrac{4}{3}T\right]
+h^2\left[ C_A^2 \left(\sfrac{23}{4}\!+\sfrac{15}{8}\xi\!
-\sfrac{1}{4}\xi^2\right)\! - 5C_A T - 4 C_F T\right]
\!+ ... ,
%+{\cal{O}}(h^3),
\\
\widetilde{\gamma}_3&=&hC_A \left(\sfrac{1}{2}+\sfrac{1}{4}\xi\right)
+h^2\left[C_A^2 \left( \sfrac{49}{24}-\sfrac{1}{16}\xi\right)
-\sfrac{5}{6}C_A T\right]
+ ... \; .
%+{\cal{O}}(h^3).
\hspace{31mm}
\end{eqnarray*}
One can easily check that
$\gamma_1-\gamma_3=\widetilde{\gamma}_1-\widetilde{\gamma}_3$
(this follows from the WST identity (\ref{WST-Z})
and the definition (\ref{an_dim})).
Moreover, since 
$\beta(g^2)=g^2\left[ 2 \widetilde{\gamma}_1
- \gamma_3- 2 \widetilde{\gamma}_3\right]$ 
(cf.\ in ref.~\cite{TVZh})
we obtain the same result for the two-loop $\beta$ function as those
given in refs.~\cite{Caswell,Jones,VlaTar,EgTar},
namely:
\be
\frac{1}{g^2}\beta\left(g^2\right)
= h \left[ -\sfrac{11}{3} C_A + \sfrac{4}{3}T \right]
+ h^2 \left[ -\sfrac{34}{3} C_A^2 + \sfrac{20}{3} C_A T 
+ 4 C_F T \right] + {\cal{O}}\left(h^3\right).
\ee 
Higher terms of the $\beta$ function are available in 
refs.~\cite{TVZh,LV,LvRV}.

%#####################################################################
\section{Conclusion} 
%#####################################################################

We have discussed the calculation of two-loop QCD vertices in 
an arbitrary covariant gauge, mainly in the zero-momentum 
limit~\cite{DOT2}. 
For the three-gluon vertex, we needed to calculate
two scalar functions, $T_1(p^2)$ and $T_2(p^2)$,
associated with different tensor structures, cf.\ eq.~(\ref{BL-decomp}).
Two independent ways of calculating these scalar functions
have been realized. One of them is just the
straightforward calculation of all diagrams 
(cf.\ Fig.~1 of ref.~\cite{DOT2}).
Another way is based on
exploiting the differential WST identity.
In this way, we obtain representations
of the scalar functions $T_1(p^2)$ and $T_2(p^2)$,
eqs.~(\ref{T1-WST}) and (\ref{T2-WST}), in terms
of the functions occurring in the ghost-gluon vertex,
its derivative (\ref{a_tilde_2}), the gluon polarization operator
and the ghost propagator. We have calculated all these
functions and confirmed the result of the straightforward
calculation.

We have constructed renormalized expressions for all Green
functions involved.
Note that in the zero-momentum limit the three-gluon vertex has no
infrared (on-shell) singularities, this is a ``pure''   
case for performing the ultraviolet renormalization.
In particular, the $Z$ factors (e.g., in the $\overline{\mbox{MS}}$
scheme) can be constructed just by using the condition
of {\em finiteness} of the renormalized Green functions. 

In principle, the techniques for calculating
scalar integrals corresponding to the two-loop QCD vertices 
in the general off-shell case
are already available~\cite{UD}.
However, some work is still needed to construct 
the complete algorithm. 

\vspace{3mm}

{\bf Acknowledgments}
A.~D. is grateful to the organizers of the Faro ERG
workshop for their hospitality.
A.~D.'s research and participation in the workshop
have been supported by the Alexander von Humboldt
foundation. 
O.~T. is grateful to BMBF for financial support   
during his stay at the University of Bielefeld.
This research has been also supported by the Research Council 
of Norway.

\end{document}